\documentclass{article}
\PassOptionsToPackage{numbers, compress}{natbib}
\usepackage[preprint]{neurips_2025}
\usepackage{fix-cm}
\usepackage{afterpage}
\usepackage{xcolor}         
\usepackage{graphicx}
\usepackage{xspace}
\definecolor{linkColor}{rgb}{0.2,0.4,0.6}
\usepackage[utf8]{inputenc} 
\usepackage[T1]{fontenc}    
\usepackage[colorlinks=true,linkcolor=linkColor,citecolor=linkColor,filecolor=linkColor,urlcolor=linkColor]{hyperref}       
\usepackage{url}            
\usepackage{booktabs}       
\usepackage{amsfonts}       
\usepackage{microtype}      

\usepackage{multirow}
\usepackage{caption}
\usepackage{amsmath}
\usepackage{float}
\usepackage{tcolorbox}
\tcbuselibrary{breakable}
\usepackage{footnotehyper} 
\makesavenoteenv{table}
\definecolor{cAzure}{HTML}{9BB1D4}
\definecolor{cGoogle}{HTML}{F2B880}
\definecolor{cGoogleZero}{HTML}{F9DFC4}
\definecolor{cOurs}{HTML}{2C5F8A}

\definecolor{abstractbg}{gray}{0.95}
\makeatletter
\renewenvironment{abstract}{%
  \vskip 0.1in
  \begin{tcolorbox}[
    colback=abstractbg, colframe=abstractbg,
    arc=3mm, boxrule=0pt,
    left=6mm, right=6mm, top=4mm, bottom=4mm,
    breakable
  ]
}{%
  \end{tcolorbox}
}
\makeatother

\usepackage{amsmath,amsfonts,bm}

\def\eqref#1{equation~\ref{#1}}

\def\1{\bm{1}}

\DeclareMathAlphabet{\mathsfit}{\encodingdefault}{\sfdefault}{m}{sl}
\SetMathAlphabet{\mathsfit}{bold}{\encodingdefault}{\sfdefault}{bx}{n}

\newcommand\our{\textsc{VibeVoice-ASR-Streaming}}
\newcommand\ourbreak{\shortstack{\textsc{VibeVoice-}\\\textsc{ASR-Streaming}}}

\newcommand{\huggingface}{\raisebox{-1.5pt}{\includegraphics[height=1.05em]{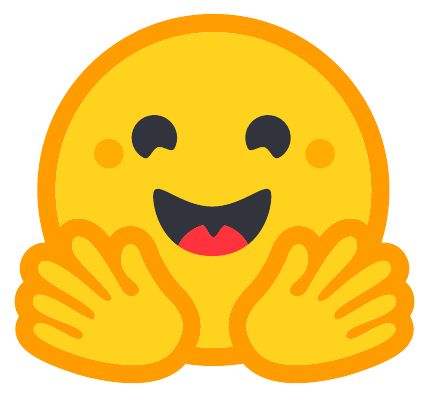}}\xspace}
\newcommand{\github}{\raisebox{-1.5pt}{\includegraphics[height=1.05em]{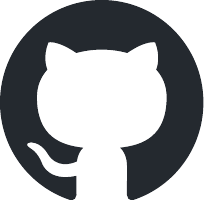}}\xspace}
\newcommand{\microphone}{{\raisebox{-1.5pt}{\includegraphics[height=1.05em]{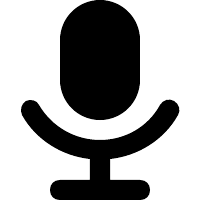}}}\xspace}

\title{\our{} Technical Report \vspace{-0.5em}
}

\author{
\bf Yujie Tu$^{2}$\thanks{Work done during Yujie Tu's internship at Microsoft Research. $\diamond$ Contact person: \href{mailto:fuwei@microsoft.com}{fuwei@microsoft.com}.}, ~~Zhiliang Peng$^{1}$, ~~Jianwei Yu$^{1}$, ~~Li Dong$^{1}$\\
\bf ~~Songchen Xu$^{3}$,~~Yaoyao Chang$^{1}$, ~~Wenhui Wang$^{1}$, ~~Zilong Wang$^{1}$\\ 
\bf ~~Zehua Wang$^{1}$,  ~~Yan Xia$^{1}$, ~~Ruibin Yuan$^{4}$,  ~~Jiajun Zhang$^{2}$,~~Xie Chen$^{3}$, ~~Furu Wei$^{1,\diamond}$ \\[4pt]
\small $^{1}$Microsoft Research \quad
\small $^{2}$University of Chinese Academy of Sciences \\ 
\small $^{3}$ Shanghai Jiao Tong University \quad \small $^{4}$ Independent Researcher \\
\small\\[2pt]
~{\href{https://aka.ms/GeneralAI}{https://aka.ms/GeneralAI}}
}

\begin{document}

\maketitle



\begin{abstract}
Traditional speaker-attributed ASR systems treated ASR and speaker diarization as two separate tasks. Recently, end-to-end models such as VibeVoice-ASR have unified the two tasks within a single model. However, existing unified models still mainly support offline recognition, making it difficult to meet the low-latency requirements of real-time voice assistants and agents. To tackle this issue, we present \our{}, one of the first LLM-based end-to-end approaches to streaming speaker-attributed ASR. It interleaves fixed-size audio chunks, a small amount of lookahead audio and previous text. This allows the model to produce ``who said what'' as speech arrives, without a separate diarization stage. For transcription accuracy, our 7B model achieves the lowest average WER/CER across five evaluation sets. For speaker attribution, it achieves the best or tied-best on 12 of 13 evaluation settings. We release the 1.5B and 7B model weights together with inference code.
\end{abstract}


\begin{table}[H]
\centering
\begin{tabular}{@{}c c@{}}
\github
\textbf{Code}: \href{https://github.com/microsoft/VibeVoice}{\texttt{github.com/microsoft/VibeVoice}} \\

\microphone
\textbf{Demo}: \href{https://aka.ms/vibeasr}{\texttt{microsoft/VibeVoice-ASR-Streaming}} \\

\huggingface
\textbf{Checkpoint}: \href{https://huggingface.co/collections/microsoft/vibevoice-68a2ef24a875c44be47b034f}{\texttt{microsoft/VibeVoice-Collection}} \\
\end{tabular}
\end{table}

\begin{figure}[!h]
\centering
\includegraphics[width=0.75\linewidth]{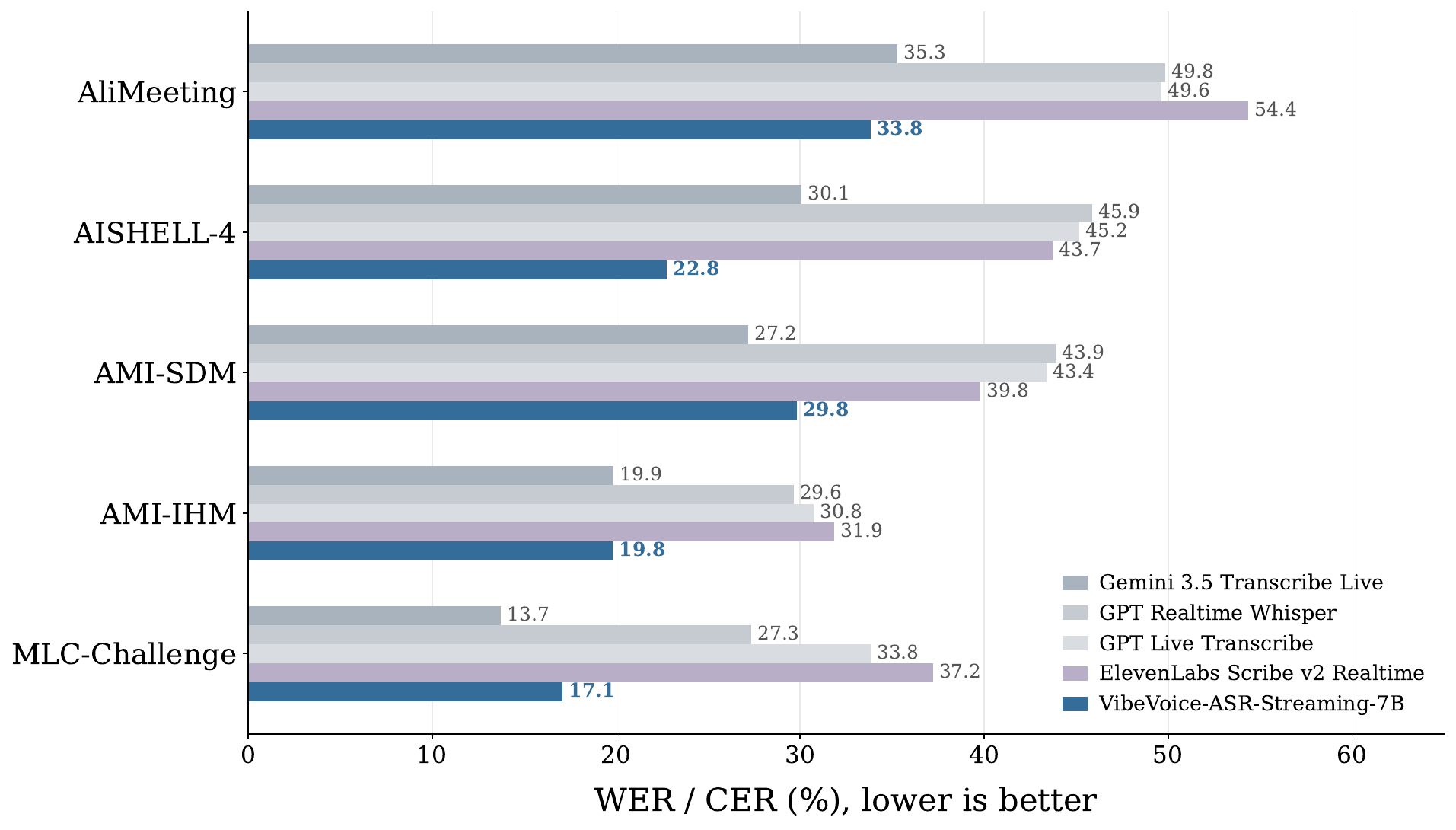}
\caption{
Recognition error of \our{}-7B and four deployed streaming ASR systems on the four meeting benchmarks and MLC-Challenge. AliMeeting and AISHELL-4 are evaluated with CER, while AMI-SDM and AMI-IHM are evaluated with WER. For MLC-Challenge, Japanese and Korean are evaluated with CER and the remaining seven languages with WER; the reported value is the macro average over the nine evaluated languages.
}
\label{fig:result}
\end{figure}


\section{Introduction}

Streaming speaker-attributed ASR must output both the words and their speaker labels as the conversation unfolds. Each sentence is attributed to a speaker when it is emitted, rather than after the recording ends.
This capability has become increasingly valuable as speech interaction has attracted growing attention in recent years.
When a voice agent is in a conversation with more than one person, it has to identify who is speaking while they are still speaking in order to process the information correctly and reduce its response latency.

Three lines of work bear on this.
LLM-based recognizers now transcribe long recordings and assign speakers in a single generative pass, including VibeVoice-ASR~\cite{peng2026vibevoiceasr}, MOSS Transcribe Diarize~\cite{yu2026moss}, SoulX-Transcriber~\cite{dai2026soulx}, and SpeakerLM~\cite{yin2025speakerlm}, but they read the whole recording before emitting output.
A second line makes LLM-based ASR streamable: BESTOW~\cite{chen2024bestow} casts inference as a read--write problem, while SpeechLLM-XL~\cite{jia2024efficientstreaming} and Uni-ASR~\cite{xia2026uniasr} consume audio in chunks and carry preceding speech-text context forward, establishing the basic recipe of incremental input with retained context --- for single-speaker transcription.
A third line makes multi-talker recognition low-latency, from SURT~\cite{lu2021surt,raj2023surt2} and t-SOT~\cite{kanda2022tsot}, which serialize overlapping talkers, to systems that attach speaker identity at low latency through token-level speaker embeddings~\cite{kanda2022tvector}, an auxiliary speaker branch~\cite{raj2024surtspk}, or online diarization cascaded with a recognizer~\cite{han2021bwedaeend,liang2024fseend,park2025sortformer,medennikov2025streamingsortformer}.
Concurrent Speech-LLM systems target the same setting~\cite{shi2025jedis,peng2026gstar}.

What the two streaming lines each leave open is the requirement speaker attribution places on retained history.
For ordinary ASR, preceding context mainly helps linguistic and acoustic modeling; speaker attribution asks more of it.
A speaker who appears in the current chunk may have first appeared several minutes earlier and must still receive the same label, so the retained history does not merely help: it is what fixes the speaker identities of the conversation.
The streaming Speech-LLM recipe naturally carries this history forward, but has so far been developed for single-speaker transcription. Streaming multi-talker systems, by contrast, usually require additional speaker-related components instead of producing speaker-attributed transcripts directly from a single model.

This report presents \our{}, which meets both requirements with one model rather than two components.
Following previous streaming Speech-LLMs~\cite{jia2024efficientstreaming,xia2026uniasr}, incoming audio and generated speaker-attributed text are interleaved,
so that future acoustic context is bounded by the chunk contract while the accumulated speech, transcription, and speaker history stays in context, and diarization never becomes a stage of its own.
Each chunk is followed by a fixed 4-frame (0.5-second) lookahead. We release 1.5B and 7B model weights for the 22-frame (2.9-second) chunk configuration, with an expected speaker-attribution latency of 2.00 seconds.
Across four meeting conditions and nine languages of MLC-Challenge, the 7B 22-frame configuration achieves the best or tied-best speaker-attributed error on 12 of 13 settings, while also attaining the best overall recognition-only mean among the compared streaming systems.

This report contributes:
\begin{itemize}
\item one of the first investigations of end-to-end LLM-based streaming speaker-attributed ASR, showing that interleaved speech-text generation can support long-form streaming recognition with strong recognition and speaker-attribution performance; we release 1.5B and 7B model weights together with inference code;

\item a thorough study of the key design choices for LLM-based speaker-attributed streaming ASR, including chunk size, lookahead, model scale, and speaker-label placement, together with detailed comparisons against the non-streaming model and deployed streaming systems, as well as serving-cost analysis over long recordings.

\end{itemize}

Section~\ref{sec:related} reviews related work.
Section~\ref{sec:method} describes the architecture and streaming formulation, the training data, and the training route.
Section~\ref{sec:results} reports the main comparison against streaming systems, Section~\ref{sec:experiments} the ablations, and Section~\ref{sec:limitations} the limitations.

\section{Related Work}
\label{sec:related}

\paragraph{Long-form speaker-attributed ASR with LLMs.}
Recent large language model (LLM)-based speech recognition systems have significantly improved long-form and multi-speaker transcription.
VibeVoice-ASR~\cite{peng2026vibevoiceasr} supports single-pass processing of up to 60 minutes of audio and jointly models transcription and speaker information within a unified generative framework.
MOSS Transcribe Diarize~\cite{yu2026moss} further extends end-to-end speaker-attributed transcription with a 128k context window and supports recordings of up to 90 minutes.
SoulX-Transcriber~\cite{dai2026soulx} improves speaker discrimination and transcription robustness through speaker-aware continuous pre-training and supervised fine-tuning, and SpeakerLM~\cite{yin2025speakerlm} unifies diarization and recognition in a multimodal LLM with a flexible speaker registration mechanism.
All of these read the whole recording before emitting output.

\paragraph{Streaming LLM-based ASR.}
Several studies have explored how LLM-based ASR can operate in a streaming manner.
BESTOW~\cite{chen2024bestow} formulates streamable Speech-LLM inference as a read--write problem.
SpeechLLM-XL~\cite{jia2024efficientstreaming} processes speech in configurable chunks and autoregressively generates the corresponding text while carrying preceding speech-text context forward.
Uni-ASR~\cite{xia2026uniasr} further develops a unified streaming and non-streaming LLM-based ASR framework with context-aware training across chunks.
These works establish the basic recipe for LLM-based streaming ASR: acoustic input is consumed incrementally, while previously accumulated context is retained for subsequent recognition.

\paragraph{Streaming multi-talker and speaker-attributed recognition.}
Streaming multi-talker recognition predates the Speech-LLM era.
SURT~\cite{lu2021surt,raj2023surt2} places an unmixing module in front of a transducer, and t-SOT~\cite{kanda2022tsot} serializes multi-talker tokens onto a single branch by emission time; in both, the output index tracks overlap and emission order rather than a speaker.
Speaker-attributed variants add the missing identity constraint through an extra component: token-level speaker embeddings decoded alongside t-SOT~\cite{kanda2022tvector}, or a speaker branch inside the transducer~\cite{raj2024surtspk}.
A parallel line keeps diarization a separate module but makes it online, from streaming EEND~\cite{han2021bwedaeend,liang2024fseend} to Sortformer~\cite{park2025sortformer} and Streaming Sortformer~\cite{medennikov2025streamingsortformer}, whose arrival-ordered speaker cache is cascaded with a streaming recognizer.
On the Speech-LLM side, JEDIS-LLM~\cite{shi2025jedis} and G-STAR~\cite{peng2026gstar} attach a speaker cache to a long-audio recognizer, though G-STAR reports chunk-wise decoding rather than a streaming deployment.

\section{Method}
\label{sec:method}

\subsection{Architecture and Streaming Formulation}
\label{sec:streaming}

\begin{figure}[!h]
\centering
\includegraphics[width=\linewidth]{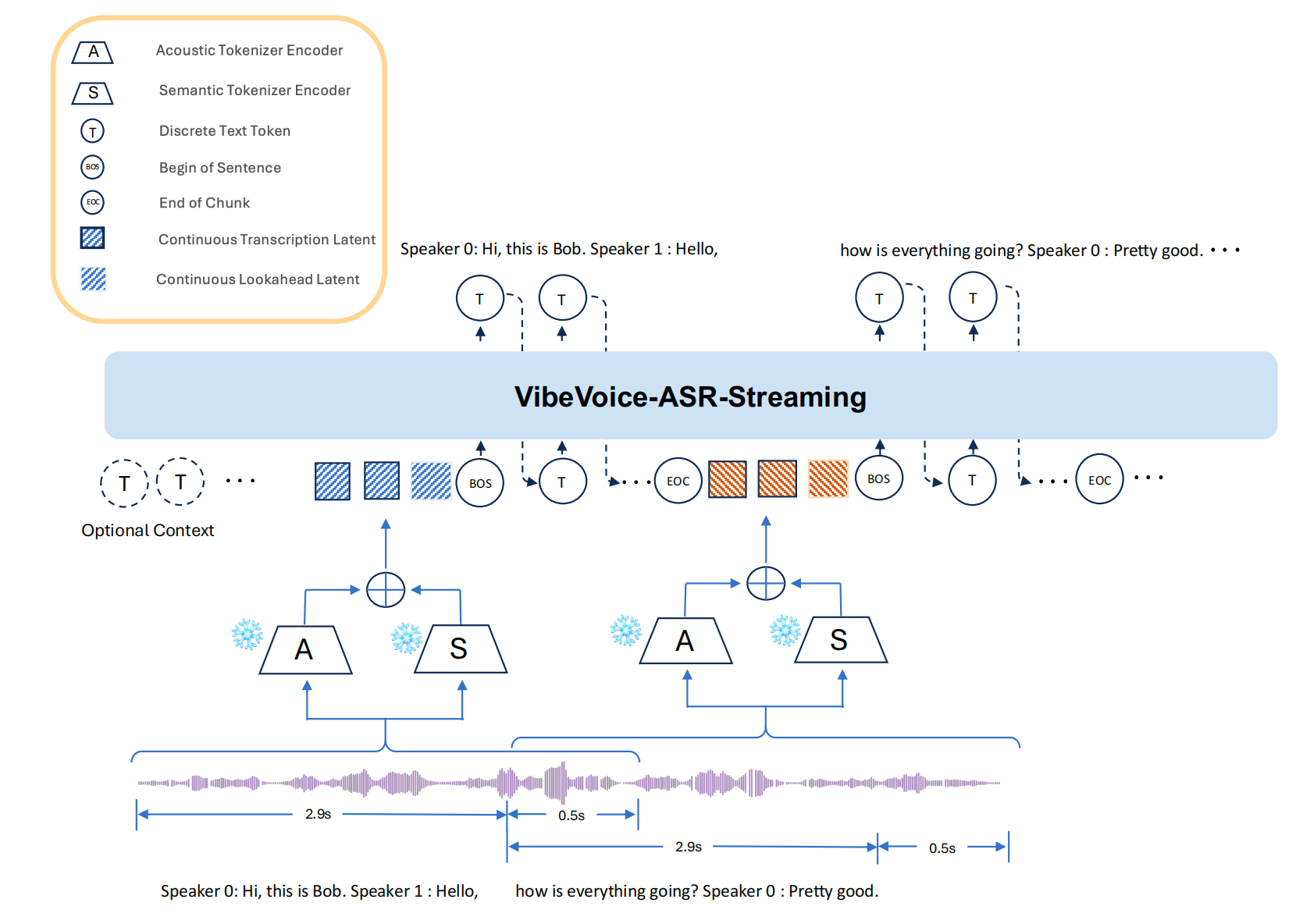}
\caption{
Architecture of \our{}.
Speech chunks $X_k$ and speaker-attributed text chunks $Y_k$ are interleaved in a single autoregressive context, and each chunk is followed by a fixed $L=4$-frame (0.5~s) lookahead before its text is generated.
}
\label{fig:overall}
\end{figure}

Figure~\ref{fig:overall} presents the architectural overview of \our{}.
Built on VibeVoice-ASR~\cite{peng2026vibevoiceasr}, \our{} extends long-form speaker-attributed transcription to streaming inference.
Speech is encoded by the pre-trained dual tokenizers of VibeVoice~\cite{vibevoice}, of which only the encoder halves are used.
The Acoustic tokenizer follows the $\sigma$-VAE design of~\cite{latentlm} and applies a hierarchical, cumulative $3{,}200\times$ downsampling to the 24-kHz waveform; the Semantic tokenizer operates at the same rate and yields deterministic features aligned with textual content.
The two therefore provide spectral detail and linguistic content on a common temporal grid. Their representations are concatenated along the feature dimension and projected into the embedding space of a Qwen2.5~\cite{qwen2_5} LLM backbone for speaker-attributed ASR.
At 24~kHz, this corresponds to one latent frame every 133.3~ms, or 7.5 frames per second. Chunk size and lookahead are therefore specified in latent frames, making every setting in this report a multiple of 133.3~ms.

Following previous streaming Speech-LLMs~\cite{jia2024efficientstreaming,xia2026uniasr}, we organize incoming speech and generated text as an interleaved sequence:
\begin{equation}
    [X_1, Y_1, X_2, Y_2, \ldots],
\end{equation}
where $X_k$ denotes the $k$-th speech chunk and $Y_k$ denotes the corresponding speaker-attributed transcription.
Unlike independent chunk-wise decoding, previously observed speech and generated text remain in the LLM context when subsequent audio arrives, so each chunk is decoded against the conversation history accumulated before it.

Retaining this history is a condition of the task rather than an optimization.
A system that discards the history has to reintroduce it elsewhere, as an external embedding store, a speaker cache, or an offline clustering pass, which reinstates the separate stage this formulation removes.

To provide limited future acoustic evidence near chunk boundaries, we introduce a fixed lookahead.
Before generating the transcription associated with each chunk, the model reads an additional $L=4$ latent frames:
\begin{equation}
    T_{\mathrm{lookahead}}
    = 4 \times \frac{3200}{24000}
    \approx 0.5~\mathrm{s}.
\end{equation}
We evaluate two chunk configurations under this lookahead: 15 latent frames, corresponding to exactly 2.0~s of audio per chunk, and 22 latent frames, corresponding to 2.9~s per chunk.

\our{} formulates ASR and speaker attribution as a single autoregressive generation task and directly produces \textit{who said what}.
After receiving the current speech chunk together with its lookahead, text generation starts as soon as the audio span is closed by the speech-end token \texttt{<|object\_ref\_end|>}.

Let $\widetilde{X}_k$ denote the current speech chunk $X_k$ together with its $L$-frame lookahead.
Formally, for the text sequence
$Y_k=(y_{k,1},\ldots,y_{k,N_k})$ associated with chunk $X_k$, we have
\begin{equation}
    p(Y_k \mid X_{<k}, \widetilde{X}_k, Y_{<k})
    =
    \prod_{j=1}^{N_k}
    p\left(
        y_{k,j}
        \mid
        X_{<k},
        \widetilde{X}_k,
        Y_{<k},
        y_{k,<j}
    \right).
\end{equation}
Here, $X_{<k}$ denotes the previously observed speech chunks, while $\widetilde{X}_k$ contains the current chunk and the $L$ future latent frames used as lookahead.
Each $Y_k$ ends with a special \texttt{<|text\_chunk\_end|>} token.
Since the input does not specify how long a chunk's transcription should be, the model must decide when to emit this token.

The token is supervised at every chunk boundary, including those with empty target text, and its emission hands control back to the audio stream.
After $Y_k$ is generated, the next speech chunk $X_{k+1}$ is appended to the same autoregressive sequence and decoding continues.

\our{} also retains the contextual prompting capability of VibeVoice-ASR~\cite{peng2026vibevoiceasr}.
Optional context, including names, technical terms, abbreviations, and other hotwords, can be provided before decoding and remains accessible throughout the streaming session.

\paragraph{Output format.}
Each $Y_k$ is a sequence of speaker-labeled utterances, so concatenating the per-chunk outputs already yields the speaker-attributed transcript.
Speakers are identified by ordinal labels assigned in order of first appearance, and a label introduced in an early chunk is reused whenever that speaker is recognized again.
Because the transcript is serialized, simultaneous speech is emitted as consecutive labeled segments rather than as parallel streams; Section~\ref{sec:limitations} discusses the consequences.
Appendix~\ref{app:output} gives the exact label syntax and a verbatim decoding trace.

Keeping the history uncompressed has a cost that grows linearly with recording length.
Inference uses the same chunk and lookahead contract as training, and the released checkpoints target recordings of up to eight minutes.

\subsection{Training Data}
\label{sec:data}

All training recordings, real and synthetic alike, are prepared the same way: word-level timing is obtained by running Qwen3-ForcedAligner-0.6B~\cite{qwen3_forced_aligner} over the recording, and the reference transcript is then split into per-chunk targets by the rule Appendix~\ref{app:output} states.

Part of the mixture is synthesized rather than collected, to improve robustness to multi-speaker acoustic conditions and specialized vocabulary.
We generate meeting-style multi-speaker conversations with domain-specific terminology and proper nouns inserted into the dialogue, keeping spoken and written forms separate: the spoken form drives speech synthesis while the written form is retained as the ASR target, so numbers, abbreviations, and technical terms are spoken naturally but transcribed canonically.
The synthesized speech then receives waveform-level augmentation: speakers are overlapped, and the mixture is convolved with room impulse responses, which apply room reverberation and microphone response in a single step.
Speaker labels and alignment are updated alongside the waveform so that the supervision survives augmentation.
This yields 50,884 recordings totaling 4,519.6 hours of augmented multi-speaker training speech.

\subsection{Training Route}
\label{sec:training}

We train \our{} in three stages that differ in how training samples are constructed rather than in the model or the training objective.

\paragraph{Stage 1: non-streaming training.}
The model is first trained in the offline speaker-attributed setting, where the complete recording is visible before the transcription is generated.
This stage establishes the basic multi-speaker recognition and speaker-attribution ability without any streaming constraint.

\paragraph{Stage 2: streaming pre-training.}
Starting from the Stage-1 checkpoint, we switch the sample construction to the interleaved form of Section~\ref{sec:streaming}: each recording is segmented into chunks, every chunk is paired with its own speaker-attributed transcription, and the fixed lookahead is appended before the corresponding text is generated.
Nothing else changes: the architecture, the set of trainable modules, and the autoregressive objective are identical to Stage 1.
The model therefore only has to adapt to bounded future context instead of relearning speaker-attributed transcription from scratch.

\paragraph{Stage 3: streaming fine-tuning.}
The streaming model is finally fine-tuned under the same interleaved formulation to obtain the reported systems.
Stage 2 draws on a subset of the Stage-1 corpus, roughly 420,000 hours of English and Chinese speech, and its job is to make the streaming format the model's normal operating condition.
Stage 3 switches to a much smaller curated mixture, about 13,000 hours drawn from public training splits and from the synthetic multi-speaker data of Section~\ref{sec:data}, and its job is to settle the behavior a user actually experiences: transcription conventions, consistent speaker labeling, and reliable hotword following.
Optimizer settings and run scale are given in Appendix~\ref{app:output}.

Each streaming configuration is initialized from the non-streaming checkpoint of the same scale, avoiding full training from scratch. The chunk size is fixed throughout Stages 2 and 3, and the 15- and 22-frame configurations are trained independently.

\section{Results}
\label{sec:results}

All models use the frozen Acoustic and Semantic tokenizer encoders and the trainable Qwen2.5 LLM backbone of Section~\ref{sec:streaming}, and differ only in backbone scale, 1.5B and 7B.
Each scale is trained at two chunk sizes, 22 latent frames (2.9~s of audio) and 15 latent frames (2.0~s), under the same fixed 4-frame (0.5~s) lookahead; unless otherwise stated the reported results use the 7B model with 22-frame chunks.
\begingroup
\setcounter{footnote}{0}
\renewcommand{\theHfootnote}{tableone-\arabic{footnote}}

\begin{table}[t]
\centering
\caption{
Recognition-only streaming ASR results on the four meeting benchmarks and MLC-Challenge. For all online APIs, audio is streamed using 2.9-s chunks, matching the chunk size of \our{}. Best results are shown in \textbf{bold} and second-best \underline{underlined}. Chinese, Japanese, and Korean are evaluated with CER, and the remaining languages with WER. The final row reports the mean over the four meeting benchmarks and the MLC-Challenge macro average.
}
\label{tab:live_recognition}
\scriptsize
\setlength{\tabcolsep}{1.2pt}
\begin{tabular}{ll|c|c|c|c|c}
\toprule
&
& \multicolumn{1}{c|}{\shortstack{
Gemini 3.5\\
Transcribe Live\footnote{\label{fn:gemini-live}\url{https://ai.google.dev/gemini-api/docs/models/gemini-3.5-transcribe}}
}}
& \multicolumn{1}{c|}{\shortstack{
GPT Realtime\\
Whisper\footnote{\label{fn:gpt-realtime-whisper}\url{https://developers.openai.com/api/docs/models/gpt-realtime-whisper}}
}}
& \multicolumn{1}{c|}{\shortstack{
GPT Live\\
Transcribe\footnote{\label{fn:gpt-live-transcribe}\url{https://developers.openai.com/api/docs/models/gpt-live-transcribe}}
}}
& \multicolumn{1}{c|}{\shortstack{
ElevenLabs\\
Scribe v2 Realtime\footnote{\label{fn:elevenlabs}\url{https://elevenlabs.io/docs/api-reference/speech-to-text/v-1-speech-to-text-realtime}}
}}
& \multicolumn{1}{c}{\shortstack{\our{}\\7B}} \\
\cmidrule(lr){3-3}
\cmidrule(lr){4-4}
\cmidrule(lr){5-5}
\cmidrule(lr){6-6}
\cmidrule(lr){7-7}

Benchmark & Language
& WER/CER & WER/CER & WER/CER & WER/CER & WER/CER \\
\midrule

\multirow{10}{*}{MLC-Challenge}
& English
& \textbf{8.40}
& 20.66
& 21.88
& 22.75
& \underline{8.44} \\

& French
& \underline{19.48}
& 27.90
& 37.83
& 32.98
& \textbf{16.42} \\

& German
& \textbf{16.96}
& 33.25
& 37.35
& 48.68
& \underline{21.83} \\

& Italian
& \textbf{12.98}
& 28.41
& 35.72
& 35.98
& \underline{17.00} \\

& Japanese
& \textbf{16.54}
& 36.50
& 33.48
& 33.17
& \underline{27.85} \\

& Korean
& \underline{9.59}
& 24.98
& 32.18
& 31.53
& \textbf{9.09} \\

& Portuguese
& \textbf{20.83}
& 38.86
& 53.46
& 59.20
& \underline{28.15} \\

& Russian
& \textbf{9.66}
& 20.55
& 29.38
& 46.78
& \underline{15.67} \\

& Spanish
& \textbf{9.13}
& 14.88
& 23.13
& 23.79
& \underline{9.31} \\

\cmidrule(lr){2-7}
& Average
& \textbf{13.73}
& 27.33
& 33.82
& 37.21
& \underline{17.09} \\

\midrule

AliMeeting & Chinese
& \underline{35.29}
& 49.84
& 49.64
& 54.35
& \textbf{33.83} \\

\midrule

AISHELL-4 & Chinese
& \underline{30.09}
& 45.86
& 45.17
& 43.74
& \textbf{22.76} \\

\midrule

AMI-SDM & English
& \textbf{27.18}
& 43.90
& 43.38
& 39.81
& \underline{29.81} \\

\midrule

AMI-IHM & English
& \underline{19.85}
& 29.64
& 30.75
& 31.86
& \textbf{19.83} \\

\midrule

Average & Mix
& \underline{25.23}
& 39.31
& 40.55
& 41.39
& \textbf{24.66} \\

\bottomrule
\end{tabular}
\end{table}

\endgroup

\begin{table}[t]
\centering
\caption{
Overall streaming speaker-attributed ASR results for the 7B model with 22-frame (2.9~s) chunks.
Chinese, Japanese, and Korean are evaluated with CER and cpCER, and the remaining languages with WER and cpWER.
For Google STT, $D_0$ uses the speaker label at first emission, while $D_{\mathrm{fin}}$ uses the final revised label after the full recording has been processed.
Best results are shown in \textbf{bold} and second-best \underline{underlined}, ranked over the systems that report the metric.
A dash marks a language or corpus the system does not support.
}
\label{tab:streaming_results}
\scriptsize
\setlength{\tabcolsep}{2.0pt}
\begin{tabular}{ll|cc|ccc|cc}
\toprule
&
& \multicolumn{2}{c|}{Azure CT\footnotemark[6]}
& \multicolumn{3}{c|}{Google STT\footnotemark[7]}
& \multicolumn{2}{c}{\shortstack{\our{}\\7B}} \\
\cmidrule(lr){3-4}
\cmidrule(lr){5-7}
\cmidrule(lr){8-9}

Dataset & Language
& WER/CER & cpWER/cpCER
& WER/CER & cpWER/cpCER & cpWER/cpCER
& WER/CER & cpWER/cpCER \\
\midrule

\multicolumn{2}{l|}{Avg.\ latency}
& \multicolumn{2}{c|}{$8.21$ s}
& --
& $9.12$ s ($D_0$)
& $51.06$ s ($D_{\mathrm{fin}}$)
& \multicolumn{2}{c}{$2.00$ s} \\
\midrule

\multirow{10}{*}{MLC-Challenge}
& English
& \underline{9.31} & \underline{23.02}
& 10.57 & 60.57 & 28.40
& \textbf{8.44} & \textbf{11.99} \\

& French
& \underline{20.55} & \underline{33.58}
& -- & -- & --
& \textbf{16.42} & \textbf{21.50} \\

& German
& \textbf{21.12} & \underline{27.17}
& -- & -- & --
& \underline{21.83} & \textbf{24.68} \\

& Italian
& \textbf{13.67} & \underline{21.53}
& -- & -- & --
& \underline{17.00} & \textbf{21.20} \\

& Japanese
& \textbf{15.04} & \textbf{33.01}
& -- & -- & --
& \underline{27.85} & \textbf{33.01} \\

& Korean
& \underline{10.32} & \underline{23.97}
& -- & -- & --
& \textbf{9.09} & \textbf{23.22} \\

& Portuguese
& \textbf{22.76} & \textbf{36.11}
& -- & -- & --
& \underline{28.15} & \underline{38.25} \\

& Russian
& \textbf{14.99} & \underline{28.11}
& -- & -- & --
& \underline{15.67} & \textbf{19.47} \\

& Spanish
& \underline{11.12} & \underline{17.00}
& -- & -- & --
& \textbf{9.31} & \textbf{11.46} \\

\cmidrule(lr){2-9}
& AVERAGE
& \textbf{15.43} & \underline{27.06}
& -- & -- & --
& \underline{17.09} & \textbf{22.75} \\
\midrule

AliMeeting & Chinese
& \textbf{29.40} & \underline{52.25}
& -- & -- & --
& \underline{33.83} & \textbf{39.80} \\
\midrule

AISHELL-4 & Chinese
& \underline{24.03} & \underline{32.51}
& -- & -- & --
& \textbf{22.76} & \textbf{28.70} \\
\midrule

AMI-SDM & English
& 33.27 & \underline{41.40}
& \underline{32.11} & 73.63 & 46.65
& \textbf{29.81} & \textbf{39.01} \\
\midrule

AMI-IHM & English
& 24.01 & \underline{31.90}
& \underline{23.07} & 71.48 & 40.15
& \textbf{19.83} & \textbf{27.48} \\
\bottomrule
\end{tabular}
\end{table}

\afterpage{%
  \footnote{%
    \url{https://learn.microsoft.com/en-us/azure/ai-services/speech-service/get-started-stt-diarization}%
  }%
  \footnote{%
    \url{https://docs.cloud.google.com/speech-to-text/docs/multiple-voices}%
  }%
}

\paragraph{Datasets.}
We evaluate on the Chinese meeting corpora AISHELL-4~\cite{fu2021aishell} and AliMeeting~\cite{yu2022m2met}, on AMI~\cite{carletta2005ami} in both its individual-headset (AMI-IHM) and single-distant-microphone (AMI-SDM) conditions, and on nine languages of the conversational benchmark MLC-Challenge~\cite{mlc}: English, French, German, Italian, Japanese, Korean, Portuguese, Russian, and Spanish. 
The benchmark itself covers more languages than these, but the forced aligner of Section~\ref{sec:data} does not, so the remaining languages are absent from training and we do not report them. All evaluation recordings are capped at 480 seconds to match the maximum session length supported by the released checkpoints.
Single-speaker results are additionally reported on AISHELL-1~\cite{bu2017aishell1}, LibriSpeech~\cite{librispeech} test-clean and test-other, and GigaSpeech~\cite{chen2021gigaspeech}.
Several of these corpora also contribute to training, but only through their official training splits; no evaluation utterance appears in any training mixture.

\paragraph{Metrics.}
We follow the MeetEval~\cite{vonneumann2023meeteval}\footnote{\url{https://github.com/fgnt/meeteval}} protocol and report word error rate (WER), which ignores speaker attribution and so reflects recognition quality alone, and concatenated minimum-permutation WER (cpWER), which concatenates the hypotheses and references belonging to each speaker and takes the minimum error over speaker permutations.
Chinese, Japanese, and Korean are scored at the character level for every system alike, as CER and cpCER, and columns headed WER and cpWER carry those values on any row or language so scored, including inside the MLC-Challenge average.
Speaker-attribution latency is the delay between a word being spoken and its speaker-attributed transcription settling: an \emph{expected algorithmic} delay $C/2+T_{\mathrm{lookahead}}$ for \our{}, with $C$ the chunk duration, giving 2.00~s at 22 frames and 1.53~s at 15, and a \emph{measured wall-clock} mean for the cloud services.
Appendix~\ref{app:metrics} gives the scoring and measurement details.

\paragraph{Compared systems.}
For recognition-only comparison, Figure~\ref{fig:result} and
Table~\ref{tab:live_recognition} additionally include
Gemini 3.5 Transcribe Live\textsuperscript{\ref{fn:gemini-live}},
GPT Realtime Whisper\textsuperscript{\ref{fn:gpt-realtime-whisper}},
GPT Live Transcribe\textsuperscript{\ref{fn:gpt-live-transcribe}},
and ElevenLabs Scribe v2 Realtime\textsuperscript{\ref{fn:elevenlabs}}.
For speaker-attributed recognition, we compare against Microsoft Azure
ConversationTranscriber (Azure CT) and Google Cloud Speech-to-Text (Google STT).
Google STT revises speaker labels retroactively, so we report two operating points:
$D_0$, using each speaker label when it is first emitted, and $D_{\mathrm{fin}}$,
using the final label after the entire recording has been processed.

\paragraph{Results.}
Figure~\ref{fig:result} summarizes recognition error on the four meeting benchmarks and the macro-averaged MLC-Challenge result, while Table~\ref{tab:live_recognition} provides the full per-language breakdown. \our{} is best on AISHELL-4, AliMeeting, and AMI-IHM, while Gemini 3.5 Transcribe Live is best on AMI-SDM. The five-set mean is 24.66 for \our{}, compared with 25.23 for Gemini 3.5 Transcribe Live, 39.31 for GPT Realtime Whisper, 40.55 for GPT Live Transcribe, and 41.39 for ElevenLabs Scribe v2 Realtime.

Across the 13 speaker-attributed settings in Table~\ref{tab:streaming_results}, \our{} gives the best or tied-best cpWER/cpCER on 12, improving over Azure CT by 2.39 to 12.45 points on the four meeting benchmarks and taking the best or tied-best value on eight of the nine MLC-Challenge languages.
It commits far earlier than the cloud services, after an expected 2.00~s against a measured 8.21~s for Azure CT and 9.12~s for Google STT, whose labels are still being revised tens of seconds later.

Single-speaker short-form audio is not what this model is built for, and it wins no individual test set in Table~\ref{tab:shortform}.
It nonetheless stays close to the strongest system on every set: with 22-frame chunks it is second on AISHELL-1 and on both LibriSpeech splits, and its four-set mean is level with the best.
The 22-frame configuration beats the 15-frame one on all four sets.

\begin{table}[h]
\centering
\caption{Single-speaker recognition, reported as a class check rather than as a competitive claim.
CER for AISHELL-1, WER for the others; the final column is the uniform mean over the four sets.
\textbf{Bold} marks the lowest value in each column and \underline{underline} the second lowest.}
\label{tab:shortform}
\small
\setlength{\tabcolsep}{2.5pt}
\begin{tabular}{l|cccc|c}
\toprule
System & AISHELL-1 & LS test-clean & LS test-other & GigaSpeech & Avg. \\
\midrule
\our{}-7B (15 frm)    & 4.81 & 2.68 & 7.40 & 10.60 & 6.37 \\
\our{}-7B (22 frm)    & \underline{4.01} & \underline{2.33} & \underline{6.49} & 10.20 & \textbf{5.76} \\
\midrule
Nemotron-3.5-ASR 0.6B~\cite{nemotron2026asr} & 13.56 & 3.02 & 6.80 & 11.76 & 8.79 \\
Voxtral-Mini-4B-Realtime~\cite{voxtral2026realtime} & 9.56 & \textbf{1.88} & \textbf{4.10} & \underline{10.11} & 6.41 \\
X-ASR~\cite{xasr2026}  & \textbf{3.61} & 2.96 & 6.85 & \textbf{9.64} & \underline{5.77} \\
\bottomrule
\end{tabular}
\end{table}

All rows are our own measurements under a single normalization.
X-ASR is evaluated in its \path{chunk-1920ms} streaming configuration, Voxtral-Mini-4B-Realtime at \path{transcription_delay_ms=2400} (2.4~s), its longest configurable delay, and Nemotron-3.5-ASR at \path{att_context_size=[56,13]} (1.12~s), in each case the released setting closest to our chunk sizes; Nemotron is additionally given the \path{zh-CN} language identifier on AISHELL-1 and \path{en-US} elsewhere, side information our own rows do not receive.

\section{Experiments}
\label{sec:experiments}

\paragraph{Cost of the streaming conversion.}
Table~\ref{tab:nonstreaming} scores \our{}-7B with 22-frame chunks against the non-streaming VibeVoice-ASR checkpoint it is initialized from, scored identically.
WER/CER rises by 0.75 to 3.53 points, cpWER/cpCER by 5.13 to 6.67 on every benchmark; Because cpWER/cpCER reflects both recognition and speaker assignment, its larger degradation than WER/CER suggests an additional loss associated with speaker attribution.

\begin{table}[h]
\centering
\caption{Cost of the streaming conversion: \our{}-7B with 22-frame chunks against the non-streaming VibeVoice-ASR checkpoint it is initialized from, scored identically.
$\Delta$ is streaming minus non-streaming, so positive values are degradations.}
\label{tab:nonstreaming}
\small
\setlength{\tabcolsep}{5pt}
\begin{tabular}{ll|cc|cc|cc}
\toprule
& & \multicolumn{2}{c|}{Non-streaming} & \multicolumn{2}{c|}{\ourbreak} & \multicolumn{2}{c}{$\Delta$} \\
\cmidrule(lr){3-4}\cmidrule(lr){5-6}\cmidrule(lr){7-8}
Dataset & Lang. & WER & cpWER & WER & cpWER & WER & cpWER \\
\midrule
AISHELL-4     & Chinese & 19.87 & 22.03 & 22.76 & 28.70 & $+2.89$ & $+6.67$ \\
AliMeeting    & Chinese & 32.15 & 34.12 & 33.83 & 39.80 & $+1.68$ & $+5.68$ \\
AMI-IHM       & English & 19.08 & 20.98 & 19.83 & 27.48 & $+0.75$ & $+6.50$ \\
AMI-SDM       & English & 26.28 & 33.88 & 29.81 & 39.01 & $+3.53$ & $+5.13$ \\
\midrule
MLC-Challenge & Mix & 14.08 & 17.62 & 17.09 & 22.75 & $+3.01$ & $+5.13$ \\
\bottomrule
\end{tabular}
\end{table}

\paragraph{Chunk size and model scale.}
Under an identical 4-frame lookahead, enlarging the chunk from 15 to 22 latent frames improves both metrics on every benchmark and at both scales (Table~\ref{tab:chunk_ablation}): on the five-set mean it is worth 1.46 WER/CER and 4.06 cpWER/cpCER at 7B, and 1.31 and 2.91 at 1.5B.
Scale acts the same way: at a fixed chunk size, moving from 1.5B to 7B lowers the mean cpWER/cpCER by 12.76 points at 22 frames and 11.61 at 15, against 4.69 and 4.54 of WER/CER.
Both factors therefore have a stronger effect on speaker attribution than on transcription --- suggesting that longer chunks and larger backbones mainly provide richer speaker evidence rather than lexical evidence.Since 15 frames reduces the expected latency from 2.00~s to 1.53~s, the two chunk settings offer different trade-offs between latency and accuracy.

\paragraph{Speaker-label placement.}
\our{} emits the speaker label before the text of the segment it belongs to, which makes the label available from the first token of a segment.
Table~\ref{tab:speaker_position} compares the two placements under an otherwise identical configuration: 7B backbone, 22-frame chunks, 4-frame lookahead, same data.
Neither metric favors the tail: the two placements land on the same mean cpWER/cpCER, 31.55 against 31.56, and the tail is no better on WER/CER either, trailing the head by 0.98 on the five-set mean.
Within the margin these measurements carry, the two placements are best read as equivalent in accuracy.

That the tail speaker lable buys nothing runs against what a cascaded system would predict.
Where attribution is a purely acoustic decision --- embed the speech, cluster the embeddings --- a label is only as reliable as the amount of speech it was estimated from, and deferring it is precisely how such a system improves.
Google STT measures the size of that effect: revising its labels once the whole recording is available lowers cpWER by $27.0$ to $32.2$ points against the label it first emits (Table~\ref{tab:streaming_results}).
An end-to-end model does not decide the same way.
It is producing the transcript at the same time, so who is speaking is settled from lexical and conversational evidence as much as from voice, and, as a human listener does, it can settle it within the two to three seconds a chunk contains rather than needing the segment to finish.

\begin{table}[t]
\centering
\caption{Effect of chunk size and model scale under an identical 4-frame lookahead, giving expected latencies of $2.00$~s at 22 frames and $1.53$~s at 15.
\textbf{Bold} marks the better chunk size for each model.}
\label{tab:chunk_ablation}
\small
\setlength{\tabcolsep}{3pt}
\begin{tabular}{ll|cc|cc|cc|cc}
\toprule
& & \multicolumn{4}{c|}{\our{}-7B} & \multicolumn{4}{c}{\our{}-1.5B} \\
\cmidrule(lr){3-6}\cmidrule(lr){7-10}
& & \multicolumn{2}{c|}{22 frm} & \multicolumn{2}{c|}{15 frm} & \multicolumn{2}{c|}{22 frm} & \multicolumn{2}{c}{15 frm} \\
Dataset & Language & WER & cpWER & WER & cpWER & WER & cpWER & WER & cpWER \\
\midrule
AISHELL-4 & Chinese & \textbf{22.76} & \textbf{28.70} & 24.62 & 32.64 & \textbf{28.48} & \textbf{39.56} & 30.33 & 42.42 \\
AliMeeting & Chinese & \textbf{33.83} & \textbf{39.80} & 34.75 & 41.66 & \textbf{39.18} & \textbf{48.87} & 39.34 & 51.69 \\
AMI-IHM & English & \textbf{19.83} & \textbf{27.48} & 20.82 & 35.40 & \textbf{22.85} & \textbf{45.15} & 23.39 & 46.48 \\
AMI-SDM & English & \textbf{29.81} & \textbf{39.01} & 31.88 & 42.49 & \textbf{34.43} & \textbf{56.83} & 35.90 & 59.49 \\
\midrule
MLC-Challenge & Mix & \textbf{17.09} & \textbf{22.75} & 18.52 & 25.85 & \textbf{21.81} & \textbf{31.15} & 24.33 & 36.01 \\
\midrule
\multicolumn{2}{l|}{Average} & \textbf{24.66} & \textbf{31.55} & 26.12 & 35.61 & \textbf{29.35} & \textbf{44.31} & 30.66 & 47.22 \\
\bottomrule
\end{tabular}
\end{table}

\begin{table}[h]
\centering
\caption{Speaker-label placement: emitting the label before the segment text (head, the released configuration) against emitting it after (tail).
Both runs are 7B with 22-frame chunks, a 4-frame lookahead, and the same data; $\Delta$ is tail minus head.}
\label{tab:speaker_position}
\small
\setlength{\tabcolsep}{4pt}
\begin{tabular}{ll|cc|cc|cc}
\toprule
& & \multicolumn{2}{c|}{Head (released)} & \multicolumn{2}{c|}{Tail} & \multicolumn{2}{c}{$\Delta$} \\
Dataset & Language & WER & cpWER & WER & cpWER & WER & cpWER \\
\midrule
AISHELL-4     & Chinese & 22.76 & 28.70 & 24.93 & 28.96 & $+2.17$ & $+0.26$ \\
AliMeeting    & Chinese & 33.83 & 39.80 & 37.72 & 40.78 & $+3.89$ & $+0.98$ \\
AMI-IHM       & English & 19.83 & 27.48 & 20.13 & 28.22 & $+0.30$ & $+0.74$ \\
AMI-SDM       & English & 29.81 & 39.01 & 30.65 & 36.52 & $+0.84$ & $-2.49$ \\
\midrule
MLC-Challenge & Mix & 17.09 & 22.75 & 14.79 & 23.32 & $-2.30$ & $+0.57$ \\
\midrule
\multicolumn{2}{l|}{Average} & 24.66 & 31.55 & 25.64 & 31.56 & $+0.98$ & $+0.01$ \\
\bottomrule
\end{tabular}
\end{table}

\paragraph{Lookahead depth.}
Lookahead (sometimes called right context) supplies no context that a later chunk would not eventually provide; it only lets the model read across a chunk boundary before committing to the text of that chunk.
Table~\ref{tab:lookahead} varies $L$ over 0, 2, and 4 frames at a fixed 22-frame chunk with everything else held constant, the $L=4$ column being the released system; both metrics improve strictly with $L$ on all five benchmarks.
The two steps cost the same 0.267~s of expected latency: averaged over the benchmarks the first pair of frames is worth 1.27 WER/CER and 1.80 cpWER/cpCER, the second pair 1.24 and 2.54, so speaker attribution is still accelerating at $L=4$ and the released configuration reads four frames rather than fewer.

\begin{table}[h]
\centering
\caption{Lookahead ablation on the 7B model at a fixed 22-frame chunk.
$L$ is the number of future latent frames read before a chunk's text is generated; expected latency is $C/2 + L \times 133.3$~ms, so each step of the sweep costs the same 0.267~s.
The $L=0$ columns have no predecessor, hence the dashes in the $\Delta$ row.}
\label{tab:lookahead}
\small
\setlength{\tabcolsep}{4pt}
\begin{tabular}{ll|cc|cc|cc}
\toprule
& & \multicolumn{2}{c|}{$L=0$ ($1.47$~s)} & \multicolumn{2}{c|}{$L=2$ ($1.73$~s)} & \multicolumn{2}{c}{$L=4$ ($2.00$~s)} \\
Dataset & Language & WER & cpWER & WER & cpWER & WER & cpWER \\
\midrule
AISHELL-4     & Chinese & 26.07 & 31.98 & 24.11 & 29.16 & 22.76 & 28.70 \\
AliMeeting    & Chinese & 37.02 & 45.02 & 35.06 & 41.07 & 33.83 & 39.80 \\
AMI-IHM       & English & 21.83 & 34.75 & 20.97 & 34.34 & 19.83 & 27.48 \\
AMI-SDM       & English & 33.39 & 44.05 & 32.09 & 43.01 & 29.81 & 39.01 \\
\midrule
MLC-Challenge & Mix & 17.60 & 23.61 & 17.31 & 22.85 & 17.09 & 22.75 \\
\midrule
\multicolumn{2}{l|}{Average} & 27.18 & 35.88 & 25.91 & 34.09 & 24.66 & 31.55 \\
\multicolumn{2}{l|}{$\Delta$ vs.\ previous column} & -- & -- & $-1.27$ & $-1.80$ & $-1.24$ & $-2.54$ \\
\bottomrule
\end{tabular}
\end{table}

\paragraph{Real-time factor.}
The expected algorithmic latency assumes a chunk is decoded before the next one is complete, so we check it where there is least room, the 7B model at 15 frames.
Decoding a chunk costs 146 to 208~ms against a 2000~ms chunk, so the real-time factor stays at or below 0.104 (Table~\ref{tab:rtf}).

\begin{table}[h]
\centering
\caption{Measured serving cost of the 7B 15-frame configuration on one A100 80GB PCIe under vLLM in bfloat16, batch size one; median of three runs on AMI-IHM audio, the two longest inputs obtained by tiling a 237~s recording.
Per-chunk compute stays well below the 2000~ms chunk at every length.}
\label{tab:rtf}
\small
\begin{tabular}{r|rrrr}
\toprule
Audio (s) & Chunks & Decode (s) & RTF & ms / chunk \\
\midrule
 30 &  15 &  2.90 & 0.097 & 193 \\
 60 &  30 &  4.83 & 0.081 & 161 \\
120 &  60 &  8.78 & 0.073 & 146 \\
237 & 119 & 20.13 & 0.085 & 169 \\
360 & 180 & 32.54 & 0.090 & 181 \\
480 & 240 & 49.94 & 0.104 & 208 \\
\bottomrule
\end{tabular}
\end{table}

\section{Conclusion and Limitations}
\label{sec:limitations}

\our{} is a streaming speaker-attributed ASR framework built on VibeVoice-ASR~\cite{peng2026vibevoiceasr}.
Interleaving incoming speech chunks with autoregressively generated speaker-attributed text produces \textit{who said what} in a single pass, without a separate diarization stage.

On recognition-only ASR, our 7B model achieves the lowest five-set mean recognition error among the compared streaming systems. The gains are more pronounced in speaker-attributed ASR: at an expected speaker-attribution latency of 2.00~s, \our{} achieves the best or tied-best cpWER/cpCER on 12 of the 13 evaluation settings, improving over Azure ConversationTranscriber by 2.39 to 12.45 points on the meeting benchmarks and from 27.06 to 22.75 on the MLC-Challenge average.

Following VibeVoice-ASR, we commit to comprehensive open-sourcing: the model weights and high-performance inference code with vLLM support are available through the links listed on the first page, together with an online demo.

\our{} has the following limitations.
\begin{itemize}
    \item \textit{Multilingual coverage:}
    Training requires word-level alignments produced by Qwen3-ForcedAligner-0.6B\cite{qwen3_forced_aligner}, so language coverage is limited by the aligner's supported languages.
    \our{} is currently trained and evaluated on ten languages; broader coverage is left to future work.

    \item \textit{Long-duration overlap:}
    \our{} handles short conversational overlaps, but performance degrades when multiple speakers overlap for long periods. This is because the decoder must serialize overlap speech into a single output stream.
    Separation-aware representations or multi-stream decoding may help.

    \item \textit{Recording length:}
    The released checkpoints support recordings of up to eight minutes.
    Longer recordings require more context and therefore more computation, so the current limit is mainly due to compute cost rather than the model architecture.

    \item \textit{First-packet latency:}
    The reported 2.00~s latency is a steady-state expectation.
    The first output requires one full chunk plus lookahead, giving 3.5~s at 22 frames and 2.5~s at 15.
    Reducing this startup delay without sacrificing the accuracy of larger chunks remains future work.
\end{itemize}

\bibliography{vibepod}
\bibliographystyle{alpha}

\clearpage
\appendix
\section{Output Format, Target Construction, and Training Details}
\label{app:output}

The speaker label is written in plain text as
\texttt{~\textbackslash n~Speaker~$k$:}; no timestamps are emitted, and a chunk
containing no speech yields an empty $Y_k$.
A word belongs to chunk $k$ if its end time falls at or before the boundary of
chunk $k$, within a 0.1~s tolerance.

The listing below shows an example output for the released 22-frame configuration,
with one chunk arriving every 2.9~s.

{\footnotesize
\begin{verbatim}
chunk 0 : " \n Speaker 0:Okay, let's start with the deployment plan."
chunk 1 : " \n Speaker 1:Sure. I think we should first verify the latency "
chunk 2 : "on the meeting set. \n Speaker 0:Agreed. We can run that today."
chunk 3 : " \n Speaker 1:Do we also need to check the multilingual results?"
chunk 4 : ""
chunk 5 : " \n Speaker 0:Yes, especially French and German."
chunk 6 : " \n Speaker 1:Okay. I'll prepare the evaluation scripts, and "
chunk 7 : "then we can compare the numbers tomorrow morning."
chunk 8 : " \n Speaker 0:Sounds good. Please also save the per-file outputs."
chunk 9 : " \n Speaker 1:Will do. \n Speaker 0:Great."
chunk 10: "Let's review everything once the runs finish."
\end{verbatim}
}

Stages 2 and 3 use AdamW ($\beta_1{=}0.9$, $\beta_2{=}0.95$,
weight decay 0.1), gradient clipping at 2.0, bfloat16, a cosine schedule
with peak learning rate $5\times10^{-5}$, and sequences packed to
8,192 tokens.
Stage 2 is a multi-node run over the 420,000-hour corpus; Stage 3 is a
500-step run on eight GPUs at a global batch of 64 sequences with
35 warmup steps, and the released checkpoints are taken at step 400.

\section{Evaluation Protocol and Baseline Systems}
\label{app:metrics}

\paragraph{Scoring.}
Chinese, Japanese, and Korean are scored at the character level, as CER and cpCER; every other language at the word level, as WER and cpWER.
All systems are scored with the same normalization and the same MeetEval cpWER implementation, with every utterance counted.

\paragraph{Latency.}
\our{} is reported as the expected algorithmic delay $C/2 + T_{\mathrm{lookahead}}$.
For the cloud services we report a measured wall-clock delay:
\begin{equation}
\label{eq:latency}
    \bar{\ell} = \frac{\sum_{j} n_j\left(\ell_j + d_j/2\right)}{\sum_{j} n_j},
\end{equation}
where $\ell_j$ is the delay between the end of the audio span covered by emitted result $j$ and the time its speaker label settles, $n_j$ and $d_j$ are its word count and duration, and $d_j/2$ is the expected position of a word inside it.
Word counts follow the scoring tokenization, and all baseline audio is pushed at no faster than $1\times$ real time.
Table~\ref{tab:conditioning} lists the side information each system receives.

\begin{table}[h]
\centering
\caption{Side information supplied to each system at inference time. \our{} receives none of it.}
\label{tab:conditioning}
\small
\begin{tabular}{lccc}
\toprule
System & Language prior & Speaker count & Hotwords / context \\
\midrule
Azure CT            & \checkmark\ explicit locale      & -- & -- \\
Google STT          & \checkmark\ \texttt{en-US} only  & \checkmark\ oracle count & -- \\
\our{}              & --                               & -- & -- \\
\bottomrule
\end{tabular}
\end{table}
\paragraph{Azure CT.}
\texttt{ConversationTranscriber} through the Python SDK \texttt{azure-cognitiveservices-speech} 1.51.1 against an \texttt{S0}-tier resource in \texttt{eastus2}.
Each file runs in its own session from 16~kHz mono 16-bit PCM pushed through a \texttt{PushAudioInputStream}, and we consume only the final \texttt{transcribed} events.
The language prior is \texttt{zh-CN} for AISHELL-4 and AliMeeting, \texttt{en-US} for AMI, and the per-utterance locale for MLC-Challenge, with \texttt{pt-PT} for Portuguese; no speaker count is supplied.
The $8.21$~s of Table~\ref{tab:streaming_results} follows Equation~\ref{eq:latency}.

\paragraph{Google STT.}
Google Cloud Speech-to-Text v1 (\path{google.cloud.speech_v1}) through the \path{StreamingRecognize} RPC, with LINEAR16 16~kHz mono audio under the \path{en-US} language code and \path{model} left unset.
Interim results, automatic punctuation, word time offsets, and speaker diarization are enabled, with \path{min_speaker_count} set to 2 and \path{max_speaker_count} to the reference speaker count of each recording.
Coverage is English only: AMI-IHM, AMI-SDM, and the English portion of MLC-Challenge, 30.80~h and 256,419 reference words in total.
$D_0$ and $D_{\mathrm{fin}}$ carry identical text and differ only in speaker attribution, and the latency reported for Google STT times the settling of speaker labels rather than of text.
Text settles after about 1.5~s, a word's speaker label keeps changing for a further 16.5--31.3~s (per-recording medians), 42.6--67.9\% of words have their label revised at least once after first emission, and within a 30~s revision budget 66.4\% (AMI-IHM), 66.8\% (AMI-SDM), and 78.4\% (MLC English) of labels agree with $D_{\mathrm{fin}}$.

\end{document}